\documentclass[traditabstract]{aa} 
\usepackage{graphicx,url}
\usepackage{natbib}
\bibpunct{(}{)}{;}{a}{}{,}

\usepackage{txfonts}
\begin{document}
   \title{A new bright eclipsing hot subdwarf binary from the ASAS and SuperWASP surveys}

   \titlerunning{A new eclipsing hot subdwarf binary from the ASAS survey}
   \authorrunning{Schaffenroth et al.}

   \author{V.~Schaffenroth \inst{1,2} \and  S.~Geier \inst{1,3} \and H.~Drechsel \inst{1} \and U.~Heber \inst{1} \and 
   P.~Wils \inst{4} \and R.~H.~\O stensen \inst{5} \and P.~F.~L.~Maxted \inst{6} \and  G.~di~Scala \inst{7}
   }

   \institute{Dr.\,Remeis-Observatory \& ECAP, Astronomical Institute, Friedrich-Alexander University Erlangen-N\"urnberg, Sternwartstr.~7, 96049
 Bamberg, Germany\\
              \email{veronika.schaffenroth@sternwarte.uni-erlangen.de}
\and Institute for Astro- and Particle Physics, University of Innsbruck, Technikerstr. 25/8, 6020 Innsbruck, Austria
		\and European Southern Observatory, Karl-Schwarzschild-Str. 2, 85748 Garching, Germany        
         \and Vereniging voor Sterrenkunde, Belgium  
         \and
            Institute of Astronomy, KU Leuven, Celestijnenlaan 200D, B-3001 Heverlee, Belgium
         \and
         	Astrophysics Group, Keele University, Staffordshire, ST5 5BG, UK
         \and 	Carnes Hill Observatory, 34 Perisher St . , Horningsea Park, NSW, Sydney Australia
             }

   \date{Received 14 December 2012 / Accepted 01 March 2013}

\abstract{We report the discovery of a bright ($m_{\rm V}=11.6$ mag) eclipsing hot subdwarf binary of spectral type B with a late main sequence companion from the All Sky Automated Survey (ASAS\,102322-3737.0). Such systems are called HW Vir stars after the prototype. The lightcurve shows a grazing eclipse and a strong reflection effect. An orbital period of $P=0.13927\,\rm d$, an inclination of $i=65.86\degr$, a mass ratio $q=0.34$, a radial velocity semiamplitude $K_1=81.0\rm\,km\,s^{-1}$, and other parameters are derived from a combined spectroscopic and photometric analysis. The short period can only be explained by a common envelope origin of the system. The atmospheric parameters ($T_{\rm eff}=28\,400$ K, $\log{g}=5.60$) are consistent with a core helium-burning star located on the extreme horizontal branch. In agreement with that we derived the most likely sdB mass to be $M_{\rm sdB}=0.46M_{\rm \sun}$, close to the canonical mass of such objects. The companion is a late M-dwarf with a mass of $M_{\rm comp}=0.16\,M_{\sun}$. ASAS\,102322-3737.0 is the third brightest of only 12 known HW Virginis systems, which makes it an ideal target for detailed spectroscopic studies and long term photometric monitoring to search for period variations, e.g. caused by a substellar companion.}
 

   \maketitle
%

\section{Introduction}
Hot subdwarf stars (sdB) are evolved, compact stars found in the disk and halo of our Galaxy. They dominate the population of faint blue stars. Especially in the context of Galaxy evolution sdBs are important because they are believed to be the dominant source for the "UV upturn phenomenon" which is observed in elliptical galaxies \citep{brown:1997,brown:2000}. Subdwarf B stars are core helium burning stars on the extreme horizontal branch (EHB). They have very thin hydrogen envelopes \citep{saffer:1994, heber:1986} which avoid hydrogen shell burning. Therefore they evolve from the EHB directly to white dwarfs \citep{dorman:1993}.\\
The formation of sdBs requires an extraordinarily high mass loss on the red-giant branch (RGB). About half of the sdB stars are found in close binaries with periods from $\sim 0.05$ to $30\rm\,d$ \citep[e.g.][]{maxted:2001,napi,cd-30,vennes}. Hence, mass transfer must play an important role in the formation of these stars. Performing binary evolution studies \citet{han:2002,han:2003} found that common-envelope evolution, resulting from dynamically unstable mass transfer near the tip of the RGB, should produce such short-period binaries ($P=0.1-10\,{\rm d}$). The most probable mass for such sdBs is 0.47 $M_{\sun}$ \citep{han:2003}, which is called the canonical mass.\\ 
Eclipsing post-common envelope binaries that consist of sdBs and late M star companions with periods of about $2-3\,{\rm hr}$ are called HW Vir systems after the prototype. Such systems are of high value because it is possible to derive the mass of the sdB as well as the mass and the nature of the companion from a combined analysis of time resolved spectra and the lightcurve. These systems are rare but can easily be recognized by the prominent reflection effect which is the only contribution of the companion to the optical light. Until now only twelve such systems are known. Interest in HW Virginis systems has been revived by the discovery of substellar companions to the prototype HW Virginis \citep{lee:2009} and HS\,0705 \citep{qian:2009} via the light travel time method (for an up-to-date census see \citet{zorotovic:2012}). For this method a long term photometric monitoring for several years is necessary.\\
ASAS\,102322-3737.0  (ASAS\,10232 for short) was discovered  as a variable star in course of the All Sky Automated Survey by \citet{pojmanski:2003}, but misclassified as a $\delta$ Scuti star. In 2007 this system was found in the ASAS Survey by P. W. and recognized as a HW Virginis system. Another lightcurve was obtained at the Carnes Hill Observatory, Sydney, in March 2008. BVRI photometry was taken with the Euler telescope on La Silla in April 2008, confirming the presence of an eclipse. In the SuperWASP survey \citep{pollacco:2006} this system was also observed in several runs. ASAS\,10232 is the third brightest HW Vir system with a $m_{\rm V}=11.6$ mag, which makes it well suited for follow-up observations.\\

\section{Observations}
\subsection{Photometry}
Observations of ASAS\,10232 were taken in the nights of the 10., 11., 12., 13., and 31.3.2008 in the B, V, and I band with a 12" LX200 GPS Schmidt-Cassegrain Telescope at the Carnes Hill Observatory in Sydney, Australia. More lightcurves in BVRI were taken with the 1.2-m Leonhard Euler Telescope at the La Silla Obervatory, Chile, in the nights of the 23. and 25.4.2008. Moreover, a lightcurve of this system was observed with the robotic SuperWASP telescope at the South African Astronomical Observatory in three runs from May 2006 to January 2009. For this observation a broad band filter (400-800 nm) was used.
\subsection{Spectroscopy}
15 medium resolution spectra (R = 3400) of ASAS\,10232 {\bf with a exposure time of  70 seconds} were obtained with the EMMI spectrograph at the ESO-NTT Telescope in La Silla, Chile, {\bf from the 11. to 14.1.2008}. These spectra cover only a small part of the optical spectrum (3900 - 4360 \r{A}) and are well suited for radial velocity work, but not for the determination of the atmospheric parameters. Hence, another 33 spectra (R $\sim 1750$, 3800 - 6540 \r{A}) were obtained on the 20. and 25.2.2010 with the GMOS spectrograph mounted at the 8.1 m Gemini South telescope at Cerro Tololo Observatory in service mode {\bf with an exposure time between 100 and 300 seconds}. The EMMI data were reduced with the MIDAS package distributed by the European Southern Observatory (ESO). The GMOS data were reduced with the PAMELA\footnote{\url{http://www2.warwick.ac.uk/fac/sci/physics/research/astro/people/marsh/software}} and MOLLY\footnotemark[1] packages.

\section{Spectroscopic analysis}
The optical EMMI and GMOS spectra give a good phase coverage of the radial velocity curve. The GMOS spectra enable us to determine the atmospheric parameters covering about half of the orbit.

\subsection{Radial velocity curve}

The radial velocities were measured by fitting a combination of Gaussians, Lorentzians and polynomials to the Balmer and helium lines of all spectra. Assuming a circular orbit sine curves were fitted to the RV data points in fine steps over a range of test periods. For each period the $\chi^{2}$ of the best fitting sine curve was determined \citep[see][]{geier:2011}. At first the EMMI- and the GMOS-datasets were fitted separately. In each case the best solution had the same orbital period as derived from the lightcurve ($P\simeq0.139\,{\rm d}$). Both datasets cover either the full phase (EMMI) or about half of the phase (GMOS). While the RV-semiamplitudes are similar, the system velocities of the two datasets show a significant shift of about $27\rm\,km\,s^{-1}$ with respect to each other. This shift is probably due to an instrumental effect. More observations are needed to solve this issue.

\begin{figure}
\centering
\includegraphics[width=1.0\linewidth]{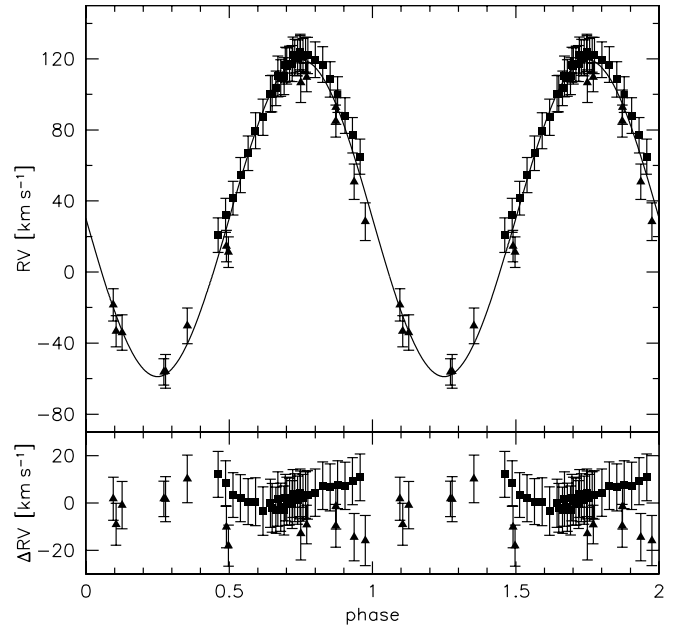}
\caption{Radial velocity plotted against orbital phase of ASAS\,10232. The RV data was folded with the period measured from the SuperWASP lightcurve. The RVs were determined from ESO-NTT/EMMI spectra (triangle) and Gemini/GMOS spectra (rectangular). The errors are formal 1 $\sigma$ uncertainties. The GMOS-RVs were shifted systematically to fit to the EMMI results.}
\label{vrad_asas}
\end{figure}

Taking the average value of the solutions derived from the two datasets the semiamplitude of the radial velocity curve was determined to $K_1=81.0\pm3 \rm\, km\, s^{-1}$. Fig.~\ref{vrad_asas} shows a phased RV-curve where the radial velocities derived from the GMOS spectra have been shifted by $-27\rm\,km\,s^{-1}$. 

\begin{table}
\caption{Radial velocities of ASAS\,10232}
\label{RVs}
\begin{center}
\begin{tabular}{lrl}
\hline
\noalign{\smallskip}
mid$-$HJD & RV [${\rm km\,s^{-1}}$] & Instrument\\
\noalign{\smallskip}
\hline
\noalign{\smallskip}
2454476.71277 &  109.7$\pm$8.2  & EMMI  \\
2454476.72692 &   85.0$\pm$9.0  &    \\
2454476.79412 &  -30.3$\pm$10.0  &    \\
2454477.71635 &   28.4$\pm$10.6  &     \\
2454477.73450 &  -33.3$\pm$8.9  &      \\
2454477.75857 &  -55.8$\pm$9.5  &      \\
2454477.84140 &   92.8$\pm$9.7  &       \\
2454477.87249 &  -18.4$\pm$9.1  &      \\
2454477.87674 &  -34.0$\pm$10.0  &      \\
2454478.66009 &  106.6$\pm$11.2  &      \\
2454478.68602 &   50.9$\pm$10.0  &      \\
2454478.76423 &   11.2$\pm$8.7  &       \\
2454478.87209 &  -56.2$\pm$7.5  &       \\
2454479.73826 &   14.6$\pm$8.9  &       \\
\noalign{\smallskip}
\hline
\noalign{\smallskip}
2455247.69381 &   127.4$\pm$10.2   &  GMOS \\
2455247.69571 &   130.4$\pm$9.8   &    \\
2455247.69801 &   136.2$\pm$10.2   &     \\
2455247.69961 &   137.3$\pm$10.4   &      \\
2455247.70092 &   143.0$\pm$10.2   &      \\
2455247.70224 &   144.3$\pm$9.9   &       \\
2455247.70355 &   144.3$\pm$10.1   &      \\
2455247.70486 &   147.9$\pm$10.1   &      \\
2455247.70618 &   147.5$\pm$10.2   &      \\
2455247.70750 &   149.9$\pm$10.2   &      \\
2455247.70881 &   148.6$\pm$9.9   &       \\
2455247.71013 &   149.3$\pm$9.8   &       \\
2455252.68221 &    47.8$\pm$9.7   &       \\
2455252.68584 &    58.9$\pm$9.6   &    \\   
2455252.68947 &    68.6$\pm$9.5   &    \\   
2455252.69310 &    81.6$\pm$9.9   &     \\   
2455252.69673 &    93.9$\pm$9.7   &      \\   
2455252.70036 &   106.6$\pm$10.2   &      \\   
2455252.70399 &   114.2$\pm$10.3   &       \\ 
2455252.70762 &   127.4$\pm$10.2   &      \\   
2455252.71125 &   137.2$\pm$11.5   &      \\   
2455252.71488 &   143.7$\pm$9.9   &      \\  
2455252.71851 &   149.1$\pm$10.1   &      \\  
2455252.72214 &   150.8$\pm$10.1   &       \\ 
2455252.72577 &   148.9$\pm$10.2   &       \\ 
2455252.72940 &   146.5$\pm$10.0   &       \\ 
2455252.73303 &   143.8$\pm$10.0   &      \\  
2455252.73666 &   135.4$\pm$10.0   &      \\  
2455252.74029 &   126.9$\pm$10.1   &      \\  
2455252.74392 &   115.1$\pm$10.0   &      \\  
2455252.74755 &   104.0$\pm$10.0   &       \\ 
2455252.75118 &    91.9$\pm$9.9   &       \\ 
2455252.75481 &    76.7$\pm$9.9   &       \\ 
\noalign{\smallskip}
\hline
\end{tabular}
\end{center}
\end{table}

\subsection{Atmospheric parameters}
\begin{figure}
\centering
\includegraphics[width=1.0\linewidth]{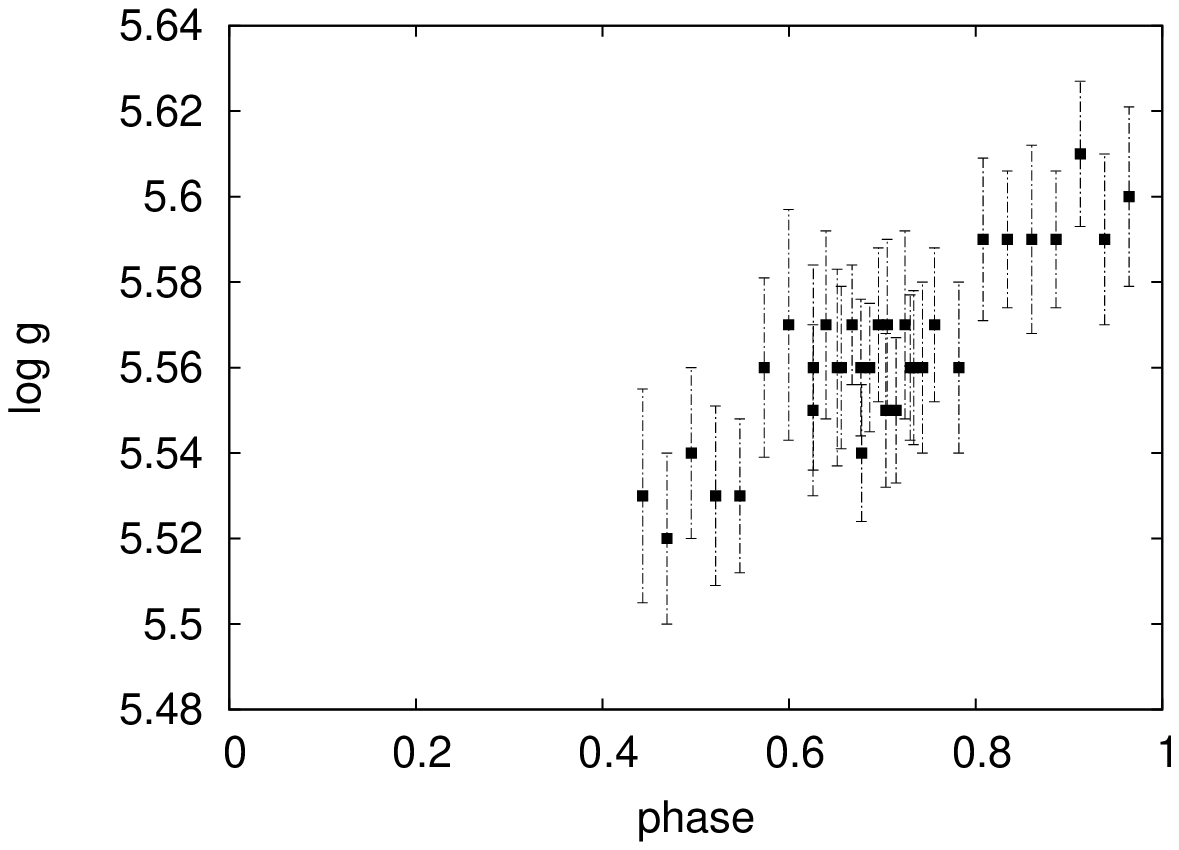}
\includegraphics[width=1.0\linewidth]{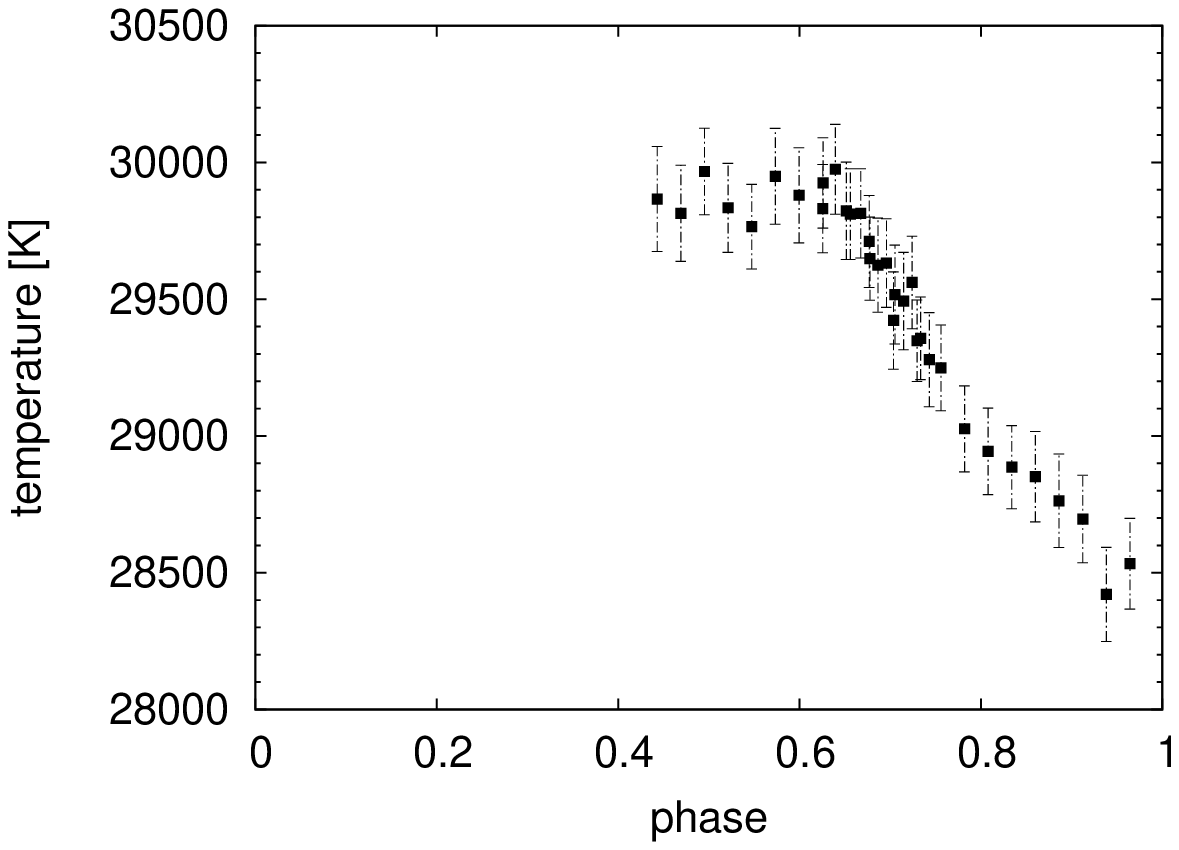}
\includegraphics[width=1.0\linewidth]{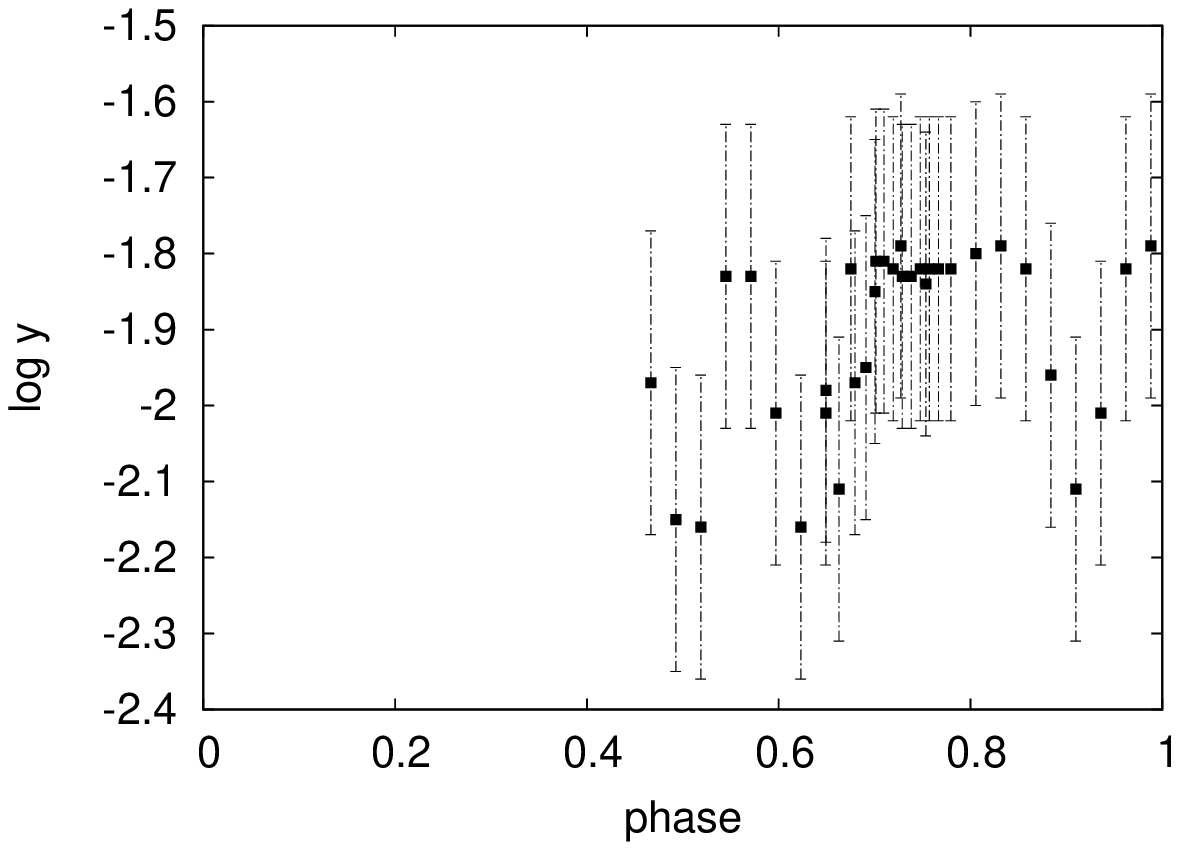}
\caption{Apparent variation of the atmospheric parameters over the phase. Temperature, surface gravity and helium abundance were measured from the Gemini/GMOS spectra. The errors are statistical errors.}
\label{logg}
\end{figure}

\begin{figure}[t!]
\begin{center}
   \includegraphics[width=\linewidth]{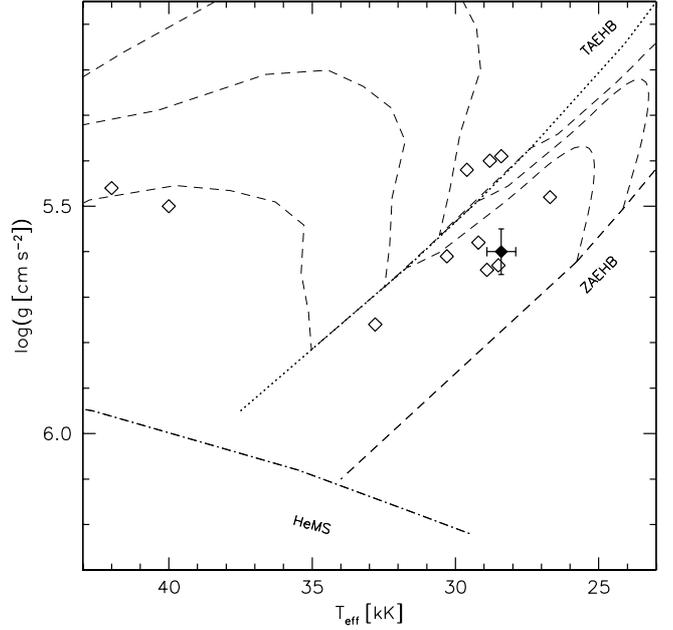}
\end{center}
\caption{$T_{\rm eff}-\log{g}$-diagram. The helium main sequence (HeMS) and the EHB band (limited by the zero-age EHB, ZAEHB, and the terminal-age EHB, TAEHB) are superimposed with EHB evolutionary tracks from \citet{dorman:1993}. The position of ASAS\,10232 is indicated with a solid diamond. Open diamonds mark the position of other HW\,Vir-like systems \citep{charpinet:2008, drechsel:2001, for:2010, geier, maxted:2002, klepp:2011, oestenson:2008, oestenson:2010, Wood:1999,almeida:2012,barlow:2012}.}
\label{tefflogg}
\end{figure}

Atmospheric parameters were determined by fitting synthetic spectra to the observed Balmer and helium lines of each of the 33 GMOS spectra using SPAS \citep{hirsch}. A model grid of synthetic spectra was calculated by using LTE model atmospheres with solar metalicity and metal line blanketing \citep{heber:2000}.
As some HW Vir systems showed an apparent change of the atmospheric parameters over the orbital phase \citep[e.g.][]{Wood:1999,drechsel:2001} all spectra were fitted separately. This effect is linked to the reflection effect as the contribution of the companion to the spectra is stronger the more the heated side is visible.\\
The variation of the parameters over the phase is clearly visible for our star system likewise. The temperature seems to change about 1500 K and the surface gravity about 0.09 dex over the phase as it can be seen in Fig.~\ref{logg}. The helium abundance shows a scatter of about 0.5 dex, which is not much more than the statistical error. For a grazing eclipse the smallest contamination by the secondary and therefore the most accurate value is expected for the phase 0. Unfortunately, the binary was not observed at this phase. We adopted the value closest to phase zero and therefore a temperature of $28400\pm 500\,\rm K$, a surface gravity of $\log{g}=5.60\pm 0.05$ and a helium abundance of about $1.8\pm 0.2$. We considered bigger errors than the statistical errors, which can be seen in Fig.~\ref{logg}, to account for the change of the parameters over the phase and systematic errors. The helium abundance is subsolar like in most sdBs. Such atmospheric parameters are typical for sdBs in HW Vir systems (see Fig.~\ref{tefflogg}).
\begin{figure}
\centering
\includegraphics[angle=-90, width=1.0\linewidth]{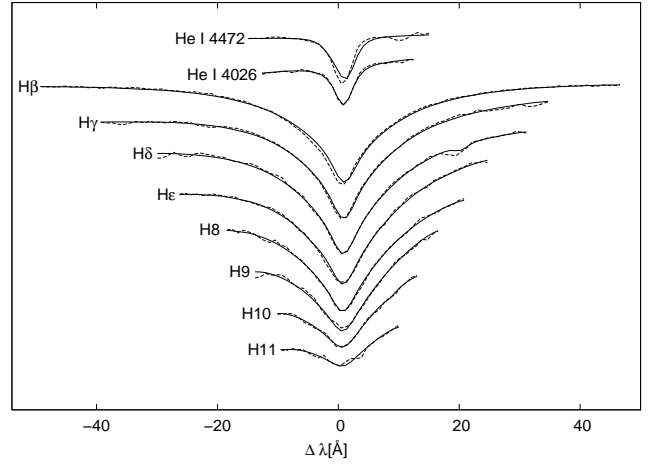}
\caption{Fit of the GMOS spectrum nearest to phase 0, the dashed line shows the measurement and the solid line shows the best fitting synthetic spectrum}
\label{spek}
\end{figure}

\section{Photometric analysis}
\begin{figure}
\centering
\includegraphics[width=1.0\linewidth]{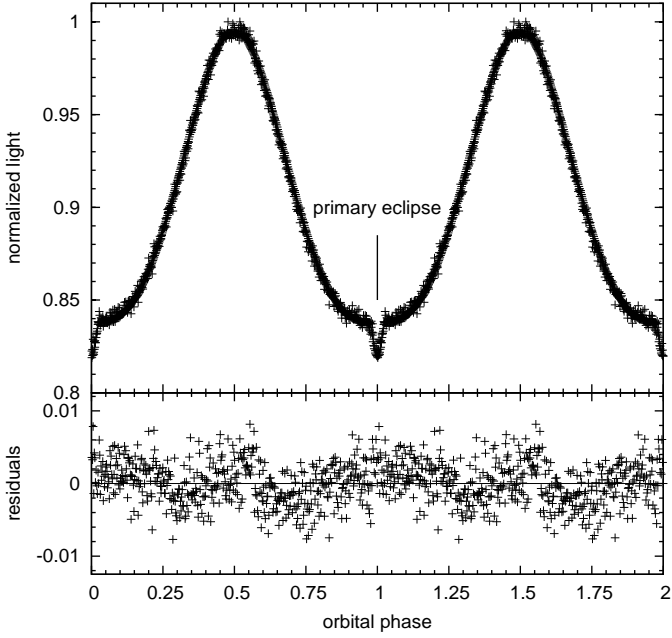}
\caption{Phased SuperWASP lightcurve. The solid line demonstrates the best fitting model. In the bottom panel the residuals can be seen. {\bf The wave pattern that is seen in the residuals can be explained by the simplified treatment of the reflection effect. Better models of the reflection effect should remove this wave pattern.} }
\label{super}
\end{figure}
\begin{figure}
\centering
\includegraphics[angle=-90,width=1.0\linewidth]{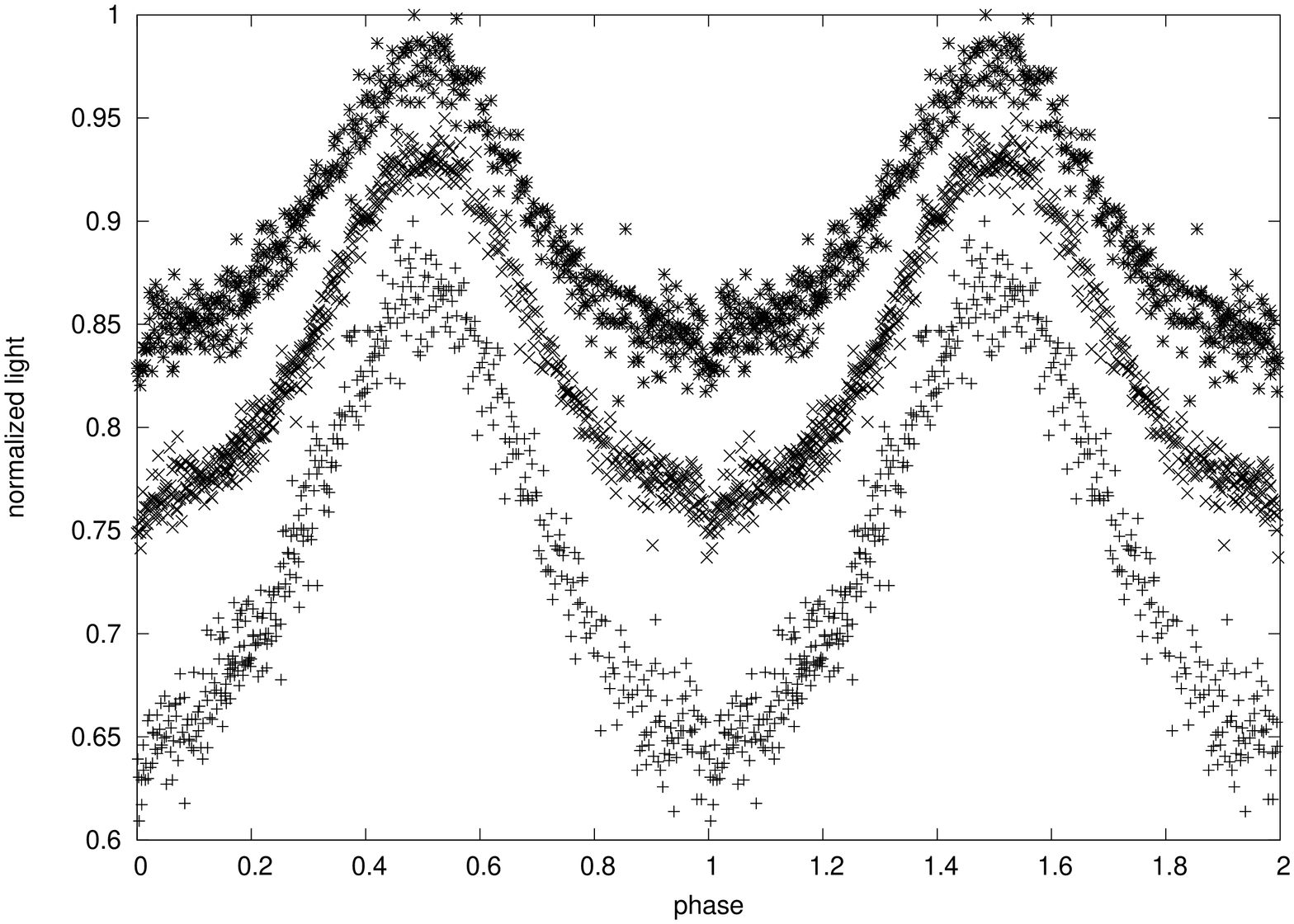}
\caption{Phased lightcurves  of ASAS\,10232 in the B ($\ast$), V ($\times$) and I ($+$) band taken from the Carnes Hill Observatory: V is shifted by 0.05, I is shifted by 0.1}
\label{asas_lc}
\end{figure}
\subsection{Ephemeris}
The SuperWASP lightcurve (see Fig.~\ref{super}) clearly shows that ASAS\,10232 is a short-period binary with a grazing eclipse. The huge reflection effect indicates a cool main sequence star companion. 
The Carnes Hill Observatory lightcurves (see Fig.~\ref{asas_lc}) in the three bands B, V and I were used to determine the time of the primary minimum. Parabolas were fitted to the minimum obtained at the night of the 13.3.2008, where the minimum was clearly visible in all bands. From the different measured times of the minima in the different bands the standard deviation was calculated as the error in time. As the SuperWASP lightcurve has much higher accuracy and covers a longer time span it was used to determine the orbital period of the system. The period derivative was found by a fit of a parabola to the O-C curve measured from the SuperWASP and BVI lightcurves, see Fig.~\ref{o-c}:
\begin{center}
HJD = $2454538\fd99689 + 0\fd13926940 \cdot E - 6\fd1 \cdot 10^{-11} \cdot E^2 $ \\
\hspace{1.1cm} $\pm 42$ \hspace{1.5cm} $\pm 4$\hspace{0.55cm} $\pm 2.3$\\
The period derivative is not yet a 3 sigma detection, more observations are necessary to confirm it.
\begin{figure}
\centering
\includegraphics[width=1.0\linewidth]{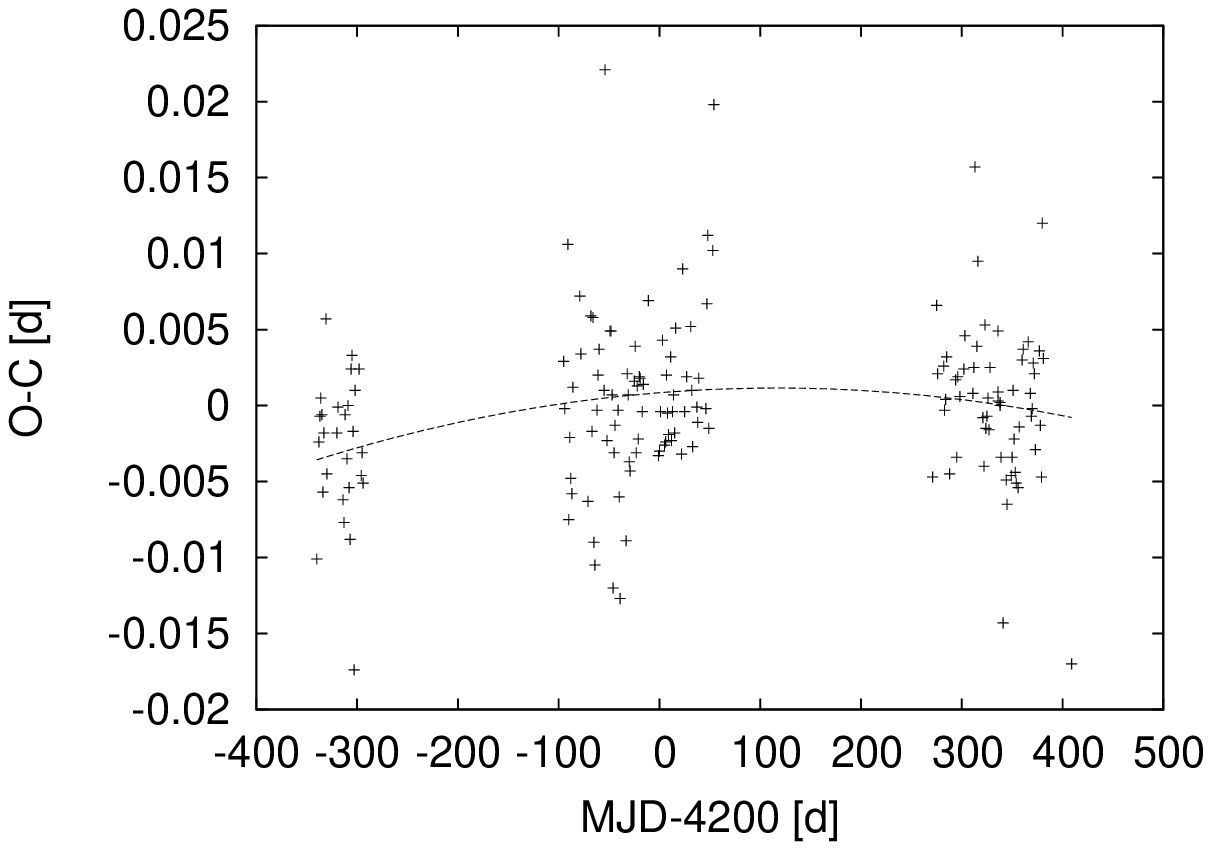}
\caption{O-C diagram of ASAS\,10232 from the SuperWasp and ASAS observations with a fit of a parabola to determine a period derivative}
\label{o-c}
\end{figure}
\end{center}
\subsection{Lightcurve solution}
\begin{figure}
\centering
\includegraphics[angle=-90,width=\linewidth]{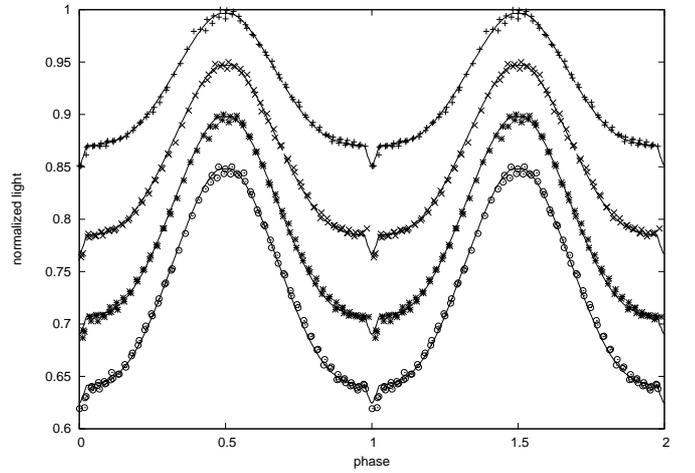}
\caption{BVRI Euler lightcurves with the model from the parameters of the best solution, but a higher $A_2=1.3$, every lightcurve is shifted by 0.05}
\label{euler}
\end{figure}

The phased B, V, I lightcurves as well as the white light SuperWASP lightcurve were analysed with the MORO code \citep{drechsel:1995}. This lightcurve solution program is based on the Wilson-Devinney approach \citep{wilson:1971}, but uses a modified Roche model that considers the radiative pressure of hot binaries. For the analysis of the SuperWASP lightcurve normal points were formed by binning the fluxes of individual measurements over narrow time intervals. This is necessary because of the huge number of data points (13816 data points were binned to 478 normal points) that would make the analysis very tedious.\\
We used the Wilson-Devinney mode 2, which does not restrict the system configuration and links the luminosity and the temperature of the second component on the basis of the Planck law. As the luminosity ratio is very high and the companion contributes almost exclusively via the reflection effect, the measured temperature of the companion is not reliable. Because of the high number of parameters (12 + 5n = 32) some have to be deducted from spectral or theoretical constraints.\\
Due to the early spectral type of the primary star the gravity darkening exponent can be fixed at $g_1=1$ as expected for radiative outer envelopes \citep{zeipel:1924}. For the cool convective companion $g_2$ was set to 0.32 \citep{lucy:1967}.  
The linear limb darkening coefficients were interpolated from the tables of \citet{wade:1985} and fixed at $x_1(B)=0.26,\, x_1(V)=0.22,\, x_1(I)=0.165,\, x_1(\rm white)=0.20$. As only monochromatic lightcurves are calculated in the lightcurve solution program we used the central wavelengths of the band filters. The secondary's limb darkening coefficient $x_2$ was varied as the limb darkening of such heated objects can not be predicted. The temperature of the hot component was taken from the spectral analysis ($T_{\rm eff,sdB}=28\,400$ K).\\ For the albedos of the companion values exceeding 1 were necessary to model the reflection effect. This can be explained with processes in the stellar atmosphere that cause a spectral redistribution of the irradiated energy with wavelength. The third light that accounts for the disturbance by a potential third object in the system was first varied but then set to 0 as it did not deviate from this value. As the spectral type of the companion is very late, radiation pressure does not play a role and the radiation pressure coefficient for the companion was fixed at zero \citep[cf.][]{drechsel:1995, drechsel:2001}.\\
The rest of the parameters such as the radiation pressure coefficient for the primary star, the inclination, the effective temperature of the companion and the Roche potentials were adjusted. A grid with different fixed mass ratios and different start parameters was calculated. All four lightcurves were analysed simultaneously. As there are so many correlated parameters a unique solution can not be found from the lightcurve alone because of the degeneracy of mass ratio and radii of the components. In Table \ref{asas} the parameters of the lightcurve solution with the best standard deviation can be found.\\ 
The Euler lightcurves are consistent with this solution as it can be seen in Fig.~\ref{euler}. The only difference is a higher albedo of 1.3 for the companion. As the Euler lightcurves have a good signal-to-noise and  four different colour bands are covered, the higher albedo is probably preferable. \\
 
\begin{figure}
\centering
\includegraphics[width=\linewidth]{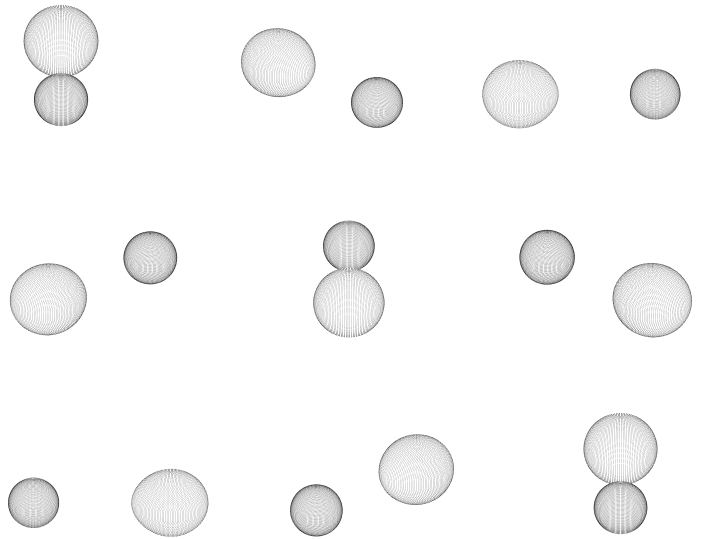}
\caption{Equipotential structure of ASAS\,10232 corresponding to the best fit solution over the phase from 0.5 to 0.5 in steps of 0.125}
\label{roche-asas}
\end{figure}

Fig.~\ref{roche-asas} shows the equipotential surfaces of both components for the case of the best solutions over the phase. It is clearly visible that ASAS\,10232 is a detached system with a grazing eclipse as both components stay inside their Roche lobe. The companion is bigger than the primary star and slightly distorted.\\


\section{System parameters}
As the radial velocity amplitude of the companion is not known and a degeneracy in the mass ratio appears in the lightcurve solutions we do not have a unique solution of the system yet. The solution with the smallest standard deviation from the lightcurve model to the measured values is found for a mass ratio $q=0.34$. For this mass ratio we derive an inclination angle of $i=65.9^\circ$ from the lightcurve analysis. Furthermore, the mean radii for the components could be determined: $R_1/a=0.185$ and $R_2/a=0.266$, where $a$ is the separation of the mass centres. Together with the $K_1=81.0\pm 3.0\, \rm km\,s^{-1}$ measured from the radial velocity curve and the mass ratio $q$, the inclination $i$ and the period $P=0.13927$ d, determined from the lightcurve, the masses of the components $M_1=0.461\pm 0.051\,M_{\sun} $ and $M_2=0.157\pm 0.017\,M_{\sun}$ were derived. Moreover, the separation $a=0.963\pm 0.036\,R_\odot$ and therefore the absolute values for the radii of the stars $R_{\rm sdB}=0.179\pm 0.011\, R_{\sun}$ and $R_{\rm comp}=0.256\pm 0.015\, R_{\sun} $ could be calculated.\\ 
Radius and mass of the companion leads to the conclusion that the companion must be an M dwarf. The derived mass for the sdB is consistent with theoretical calculations for the formation of sdBs via the common envelope channel \citep{han:2002,han:2003}.\\ In order to check this solution the surface gravity determined from the spectra can be compared to a photometric surface gravity. This surface gravity is derived via the mass - radius relation from the mass and radius calculated from the radial velocity curve together with the lightcurve analysis. The surface gravity determined in this way $\log{g}(\rm phot)=5.60\pm 0.02$ is in perfect agreement to the spectroscopic surface gravity $\log{g}(\rm spec)=5.60\pm 0.05$.
\begin{table}
\caption{Parameters of ASAS\,10232 ({\bf TYC\,7709-376-1})}
\label{mass}
\centering
\begin{tabular}{l||c|c|c}
	\hline
		coordinates	&\multicolumn{3}{c}{10 23 22\, -37 37 00 (J2000.0)}\\
		{\bf proper motions}&\multicolumn{3}{c}{{\bf -28.9 -22.8 [2.4 2.4 35] mas/yr}}\\
		\hline
		mass of the sdB & $M_{\rm sdB}$ & [$M_{\rm \sun}$] & $0.461\pm0.051$\\
		mass of the companion & $M_{\rm comp}$ & [$M_{\rm \sun}$] & $0.157\pm0.017$\\
		separation & $a$ & [$R_{\rm \sun}$] & $0.963\pm 0.036$\\
		mean radius of the sdB&$R_{\rm sdB}$ & [$R_{\rm \sun}$]& $0.179\pm 0.011$\\
		mean radius of the comp&$R_{\rm comp}$ & [$R_{\rm \sun}$]&$0.256\pm 0.015$\\
		surface gravity (phot)& $\log{g}(\rm sdB)$ & & $5.60\pm0.02$\\
		surface gravity (spec)& $\log{g}(\rm sdB)$ & & $5.60\pm0.05$\\
	\hline
\end{tabular}
\end{table}

\section{Conclusions}
We discovered the new eclipsing sdB star ASAS\,102322-3737.0, which shows a grazing eclipse and a huge reflection effect that is the only contribution of the companion to the optical light of the system. It has an orbital period of $P=0.13927$ d and an inclination of $i=65.9^\circ$. A unique solution can not be found as there exists a degeneracy in the mass ratio. The best lightcurve solution was calculated for a mass ratio of $q=0.34$. The masses of the components for this solution $M_{\rm sdB}=0.461\pm0.051\,M_{\rm \sun}$ and $M_{\rm comp}=0.157\pm0.017\,M_{\rm \sun}$ are typical for an sdB and a late main sequence star. Also the spectroscopic and the photometric surface gravity are in agreement. 
As ASAS\,10232 is very bright it is possible 
to apply rapid high resolution time series spectroscopy to search for spectral features and the RV semiamplitude  from the companion to resolve the degeneracy in the mass ratio, as done for AA Dor \citep{aador}. \\
Some of the sdBs were found to be pulsating. There are two classes of sdB pulsators, p-mode pulsators with periods of some minutes and g-mode pulsators with periods of 0.5 to 2 hours, which are separated by their atmospheric parameters. Until now only one pulsating HW Virginis system was found, NY Virginis \citep{nyvir}. Such systems are very interesting as it is possible to compare the results from asteroseismology to the results from the lightcurve analysis of the eclipses. If it is pulsating, the parameters of ASAS\,10232 would suggest it to be most likely a g-mode pulsator. However, the Euler lightcurve with the best signal-to-noise shows no sign of pulsation. But the strong reflection effect could hide the pulsations.\\
Period variations have been found for almost all of the HW Vir systems that have accurate eclipse timings covering more than five years, which may be attributed to the presence of a third body. In several cases the third bodies are likely to be one (or two)  giant planet(s). These discoveries were unexpected, because it is regarded difficult for giant planets to form  around main sequence binaries due to the short life time of circumbinary disks. In addition, such planets may not be able to survive common envelope (CE) evolution.
Instead, it has been suggested that these circumbinary planets are of second generation \citep{zorotovic:2012}, that is formed from the instability of a post-CE disk. \citet{zorotovic:2012} proposed also an alternative scenario for the period variations due to processes acting in deeply convective secondary stars. 

HW Vir has been monitored for more than 28 years now and \citet{lee:2009} found two sinusoidal variations of the light-travel time for HW Vir from 24 years of data, suggesting the presence of two substellar objects orbiting the close binary. New observations by \citet{beuermann:2012}, however, deviate significantly from the prediction of \citet{lee:2009}. The new solution involves one planet and a brown dwarf or low-mass star  orbiting the HW Vir binary \citep{beuermann:2012}. The new solution was found to be stable in contrast to that of \citet{lee:2009}.

These findings suggest that the probablity to find period variations in ASAS 10232 as well is high, whether due to substellar objects or the active secondary. The lesson learned from HW Vir is that long term and dense monitoring
is a prerequisite (Beuermann et al. 2012). ASAS is the third brightest HW Vir system known, only one magnitude fainter than HW Vir, which allows us to use readily accessible small telescopes to obtain lightcurves. Hence it is a promising target for amateur and highschool observatories to team up with professional astronomers \citep{Backhaus:2012}. 
   
\begin{table}
\caption{Best lightcurve solution of ASAS\,10232}
\label{asas}
\begin{tabular}{lcl}
\hline
\noalign{\smallskip}
\multicolumn{3}{l}{Fixed parameters:}\\
\noalign{\smallskip}
\hline
\noalign{\smallskip}
$q\,(=M_{2}/M_{1})$ & & $0.34$\\
$T_{\rm eff}(1)$&[K]&\multicolumn{1}{l}{28400}\\
$g_1^b$&&\multicolumn{1}{l}{1.0}\\
$g_2^b$&&\multicolumn{1}{l}{0.32}\\
$x_1(B)^c$&&\multicolumn{1}{l}{0.26}\\
$x_1(V)^c$&&\multicolumn{1}{l}{0.22}\\
$x_1(I)^c$&&\multicolumn{1}{l}{0.165}\\
$x_1(\rm white)^c$&&\multicolumn{1}{l}{0.20}\\
$\delta_2^d$&&\multicolumn{1}{l}{0.0}\\
$l_3^f$&&\multicolumn{1}{l}{0.0}\\
\noalign{\smallskip}
\hline
\noalign{\smallskip}
\multicolumn{3}{l}{Adjusted parameters:}\\
\noalign{\smallskip}
\hline
\noalign{\smallskip}
$i$ & [$^{\rm \circ}$] & $65.86\pm 0.69$ \\
$T_{\rm eff}(2)$ & [K]& $3380 \pm 561$\\
$A_1^a$&&\multicolumn{1}{l}{$0.94\pm0.03$}\\
$A_2^a$ & & $1.21\pm0.13$\\
$\Omega_1^f$&&$5.700 \pm 0.260$\\
$\Omega_2^f$&&$2.673 \pm 0.052$\\
$\frac{L_1}{L_1+L_2}(B)^g$&&$0.99954 \pm 0.00077 $\\
$\frac{L_1}{L_1+L_2}(V)^g$&&$0.99804 \pm 0.00247 $\\
$\frac{L_1}{L_1+L_2}(I)^g$&&$0.99036 \pm 0.00845 $\\
$\frac{L_1}{L_1+L_2}(\rm white)^g$&&$0.99709 \pm 0.00337 $\\
$\delta_1$&&$0.0123\pm 0.0080$\\
$x_2(B)$&&$0.638 \pm 0.097$\\
$x_2(V)$&&$0.548 \pm 0.0630$\\
$x_2(I)$&&$0.625 \pm 0.087$\\
$x_2(\rm white)$&&$0.238 \pm 0.089$\\
\noalign{\smallskip}
\hline
\noalign{\smallskip}
\multicolumn{3}{l}{Roche radii$^h$:}\\
\noalign{\smallskip}
\hline
\noalign{\smallskip}
$r_1$(pole)&[a]&$0.184 \pm 0.009 $\\
$r_1$(point)&[a]&$0.186 \pm 0.00 $\\
$r_1$(side)&[a]&$0.185 \pm 0.009 $\\
$r_1$(back)&[a]&$0.186 \pm 0.009 $\\
\noalign{\smallskip}
$r_2$(pole)&[a]&$0.248 \pm 0.009$\\
$r_2$(point)&[a]&$0.286 \pm 0.017 $\\
$r_2$(side)&[a]&$0.256 \pm 0.010 $\\
$r_2$(back)&[a]&$0.277 \pm 0.014 $\\
\noalign{\smallskip}
\hline
\end{tabular}\\
\tablefoot{\\
$^{a}$ Bolometric albedo\\
$^{b}$ Gravitational darkening exponent\\
$^{c}$ Linear limb darkening coefficient; taken from \citet{wade:1985} \\
$^{d}$ Radiation pressure parameter, see \citet{drechsel:1995}\\
$^{e}$ Fraction of third light at maximum\\
$^{f}$ Roche potentials\\
$^{g}$ Relative luminosity; $L_2$ is not independently adjusted, but recomputed from $r_2$ and $T_{\rm eff}$(2)\\
$^{h}$ Fractional Roche radii in units of separation of mass centers}
\end{table}   

\begin{acknowledgements}
Based on observations at the La Silla Observatory of the 
European Southern Observatory for programmes number 080.D-0685(A) and on observations obtained at the Gemini Observatory (Program ID: GS-2009B-Q-98), which is operated by the 
Association of Universities for Research in Astronomy, Inc., under a cooperative agreement 
with the NSF on behalf of the Gemini partnership: the National Science Foundation (United 
States), the Science and Technology Facilities Council (United Kingdom), the 
National Research Council (Canada), CONICYT (Chile), the Australian Research Council (Australia), 
Minist\'{e}rio da Ci\^{e}ncia e Tecnologia (Brazil) 
and Ministerio de Ciencia, Tecnolog\'{i}a e Innovaci\'{o}n Productiva (Argentina).
S.G. is supported by the Deutsche Forschungsgemeinschaft (DFG) through grant HE1356/49-1. R.\O. acknowledges funding from the European Research Council under the European Community's Seventh Framework Programme (FP7/2007--2013)/ERC grant agreement N$^{\underline{\mathrm o}}$\,227224 ({\sc prosperity}), as well as from the Research Council of K.U.Leuven grant agreement GOA/2008/04.
\end{acknowledgements}

\bibliographystyle{aa}

\end{document}